\documentclass[twocolumn,superscriptaddress,amsmath,amssymb,floatfix,aps,notitlepage,nokeys,pre]{revtex4}
\usepackage{physics}
\usepackage{mathrsfs, hyperref}
\usepackage{amssymb,amsbsy,amsmath,latexsym,dsfont,array,layout,graphics,mathrsfs,braket,amsfonts,amsthm,amssymb,graphicx,subfigure,dcolumn,youngtab,color,bm,mathtools,braket,verbatim,url,cleveref,natbib,hypernat,extarrows,inputenc,multirow,soul}
\usepackage[mathlines]{lineno}

\begin{document}
\title{Non-local twist sequences in floppy kagome chains}

\author{Pegah Azizi}
\author{ Stefano Gonella}%
\email{sgonella@umn.edu}
\affiliation{%
Department of Civil, Environmental, and Geo- Engineering,
University of Minnesota, Minneapolis, Minnesota 55455, USA}%
\date{Accepted 13 August 2024; published 3 September 2024; updated on arXiv 28 January 2025}
\text{Published DOI:\href{https://journals.aps.org/prapplied/abstract/10.1103/PhysRevApplied.22.034001}{10.1103/PhysRevApplied.22.034001}}

\begin{abstract}
The twisted kagome family comprises a spectrum of configurations, that can be realized through the sweep of a single configurational degree of freedom known as twist angle. Recently, it has been shown that certain pairs of configurations along this sweep are linked by duality transformations and display matching phonon spectra. In this work, we introduce an inter-cell connection system that spreads the lattice in the dimension orthogonal to the tessellation plane. The resulting 3D character of the lattice allows us to sweep the entirety of the twist angle spectrum, including all the compact configurations featuring overlapping triangles that, in a strictly 2D space, are forbidden. Duality provides precious guidance in interpreting the availability of floppy mechanisms arising in the compact configurations through the one-to-one correspondence with their expanded counterparts. Our focus is on the compact configuration corresponding to a null twist angle, where the lattice degenerates to a chain. From the perspective of the chain, several of the local connections between neighboring lattice cells play the role of non-local long-range interactions between cells of the chain. We demonstrate experimentally some peculiar behavior that results from such non-locality, including a selective activation of floppy sequences that is informed by the direction of loading.
\end{abstract}

\maketitle
\section{Introduction}
Mechanical metamaterials are structural material systems deriving their properties and functionalities primarily from structural design rather than chemical composition~\cite{Lee_et_all_mech_mat_AdvMat_2012,Babaee_et_sll_metamaterial_2013,Christensen_et_ell_MechMat_2015,Rafsanjani_Pasini__metamaterials_2016,metamaterials_Bertoldi2017,Deng_et_all_soliton_PRL2017,Bolei_et_all_floppy_PNAS2020}. Among them, lattice metamaterials may consist of periodic networks of springs and lumped masses, trusses of rods connected by ideal hinges, or frames of beams~\cite{brillouin_wave_1953,Phani_et_all_Acoustical_Soc_America_2006,mechanism_metamaterials_Coulais_Natcom2022}. Their stability depends on the coordination number $z$, representing the average number of connections available at each node. Lattices are classified in D-dimensional space as under- $(z<2D)$, over- $(z>2D)$, and critically coordinated $(z=2D)$, the latter case referred to as Maxwell lattices~\cite{Maxwell_1864,calladine_lattice_IJSS_1978}. Under-coordinated lattices accommodate deformation modes with zero elastic energy, termed \textit{floppy modes}, wherein stresses are not stored in the bonds. On the other hand, over-coordinated systems lack floppy modes aside from rigid-body motion~\cite{Thorpe_floppy_1983,Connelly2005}. Maxwell lattices feature the richest behavior, whereby they may support floppy modes, either in the bulk or on the edges, according to their specific cell geometry and boundary conditions~\cite{Lubensky_et_all_2015,Mao_Lubensky_APS_2011,Kane_Lubensky_Nphys_2014,Mao_Lubensky_maxwell_topo_2018}.

Among 2D Maxwell lattices, kagome lattices feature a unit cell comprised of two triangles connected at a vertex~\cite{Schaeffer_Ruzzene_kagome_appliedphys_2015,Bertoldi_et_all_natrevmats_2017,Chen_et_all_kagome_PhysRevB_2018,Riva_et_all_topo_kagome_appliedphys_2018,Nassar-et-al_Microtwist_JMPS_2020}. The kinematics of the kagome family are described by the relative twist angle $\theta$. For instance, rotating two equilateral triangles by $180^{\circ}$ yields the regular kagome lattice, which features three directions—relatively rotated by $60^\circ$—along which the bonds align perfectly, forming continuous load-carrying ``fibers'' that support states of self-stress (SSS). The SSS directions also support bulk floppy modes, which, to the leading order approximation, involve alternating rotations of the bonds located along the fibers. These bulk floppy modes appear in the lattice band diagram as a flat zero-frequency branch for wave vectors along the $\Gamma-M$ direction of the irreducible Brillouin zone (BZ) contour~\cite{Anton_et_all_kagome_prl_2009}. In contrast, alternative angles result in twisted kagome configurations that lack SSS, lifting the phonon spectrum to finite frequencies for all wave vectors—except at the origin~\cite{Sun_et_all_pnas_2012}. In practice, the twist angle sweep corresponds to a global nonlinear soft mode, known as Guest-Hutchinson mode, which deploys the lattice progressively from a compact to an expanded state~\cite{GUEST_Hutchinson_2003,Rocklin_et_all_natcom_2017,Li_Kohn_GHkagome_JMPS2023}.

\begin{figure}[t!]
\includegraphics[width=\columnwidth]{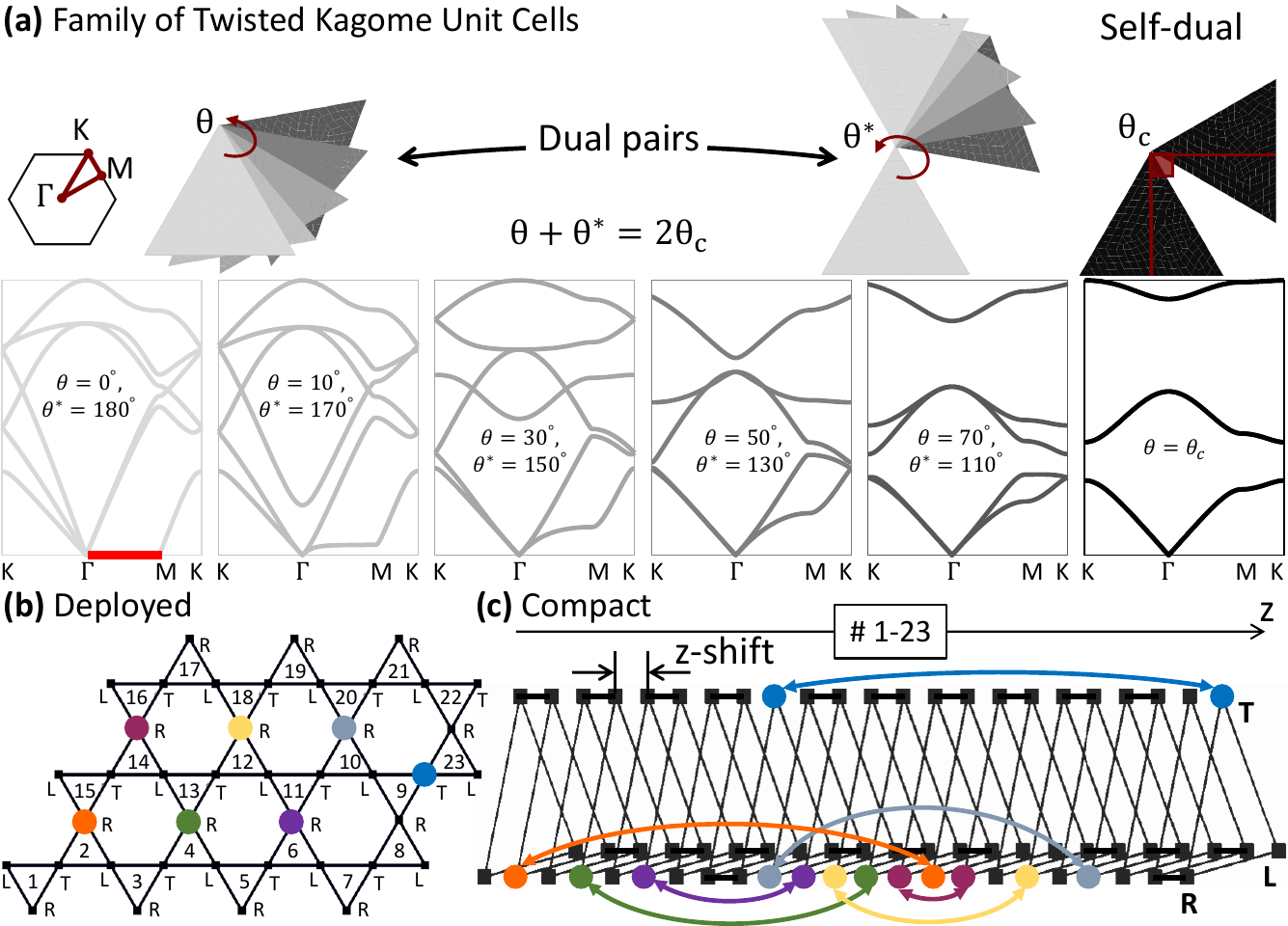}
\caption{\label{fig:model}(a) Sweep of twist angles highlighting dual pairs ($\theta$ and $\theta^\ast$) featuring identical band diagrams (color-coded). The bulk floppy mode for the dual set ($\theta = 0^\circ, 180^\circ$) is highlighted in red. The self-dual configuration ($\theta_c = 90^\circ$) is reported on the right. (b-c) Schematic representations of a finite kagome domain comprising 23 triangles, shown in the deployed state for $\theta=180^\circ$ (b) and in the compact state collapsing to a twist chain for $\theta=0^\circ$ (c). The vertices are denoted as T, L, and R and the hinges connecting the rows of cells are color-coded to illustrate their migration along the chain axis.}
\end{figure}

Recently, Fruchart \textit{et al.} discovered a special form of duality within the twisted kagome family, which can be ascribed to a hidden non-spatial symmetry~\cite{Fruchart_et_all_duality_nat_2020}. This duality links pairs of distinct configurations that are equidistant in configuration space (along the sweep axis) from a critical $\theta_c$ value. Specifically, for a given configuration with twist angle $\theta$, there exists a dual with angle $\theta^\ast$, such that $\theta+\theta^\ast=2\theta_c$, featuring an identical band diagram~\cite{Fruchart_et_all_duality_nat_2020}. The critical configuration separating the dual sets, with $\theta_c=90^{\circ}$, maps to itself and is referred to as self-dual. Despite sharing the same space group symmetry with other pairs, the self-dual configuration showcases unique properties such as double degeneracy for all phonons across the entire BZ~\cite{Azizi_et_all_PhysRevLett_2023,Fruchart_et_all_duality_nat_2020,Gonella_duality_PhysRevB_2020}. Duality is observed in ideal lattices, such as spring-mass or truss-like lattices~\cite{Gonella_duality_PhysRevB_2020,Duality_Lei_et_all_PRL2022,Duality_Fruchart_et_all_PRResearch2023}. The phenomenon is summarized in Fig.~\ref{fig:model}(a) for the case of solid elastic triangles connected by perfect hinges, where the triangles can deform internally according to 2D plane-stress elasticity while maintaining the ability to rotate freely, thus enjoying the hinge ideality conditions required by duality. Indeed we can see that dual pairs exhibit overlapping spectra.

The motivation of this study germinates from an attempt to explore holistically the \textit{entire} twisted kagome family, delving into potentially unconventional phenomena that emerge from its collapse into compact states, interpreting the mechanics of any emergent configurations through the lens of duality. Recall that, as we sweep the twist angle, the lattice transitions from a deployed state $(\theta = 180^{\circ})$ to a compact state $(\theta = 0^{\circ})$. Our investigation focuses on this extreme dual pair. Notably, duality ensures that both the $180^{\circ}$ and $0^{\circ}$ configurations feature floppy bulk modes along the high-symmetry line $\Gamma-M$ (SM.\ref{sec.1}). While the manifestation of such modes in the deployed lattice is well understood, one may ask how the phenomenon manifests in the compact case, and whether, in such case, the kinematics of the bulk floppy mode can give rise to new structural logic functionalities. 

\section{Collapsing 2D kagome lattices into 1D kagome chains}
To address this question, we propose a variation of the conventional twisted kagome specifically designed to elucidate the mechanics of the compact configuration. This task immediately meets a practical challenge: while, in principle, all values of $\theta$ are allowed, practical constraints require that $\theta$ not exceed values at which two triangles come into contact, i.e., the $\theta < 60^\circ$ and $\theta > 300^\circ$ ranges are forbidden. This restriction becomes more severe in structural lattices~\cite{Azizi_et_all_PhysRevLett_2023,Zhang_et_all_kagome_2023_PhysRevApplied,Azizi_2024_DW_PhysRevB}, where the finite thickness of the beam-like bonds and of the filleted joints at the internal clamps prevent tight re-entrant angles even in the permissible $\theta$ range. To circumvent this limitation, we propose a modified kagome arrangement in which we let pairs of triangles that exchange forces at a shared hinge be shifted by a finite amount $d$ in the out-of-plane $z$ direction. This modification enables them to rotate \textit{freely} on \textit{parallel} $z$-shifted planes. The \textit{co-planar} mechanics of the system follow the same elasticity rules of the in-plane mechanics of the original kagome lattice, albeit featuring an additional dependence on the $z$ coordinate. For the sake of illustration, let us assume the deployed state of a finite domain consisting of $n \times m$ unit cells, numbered according to a semi-row-wise pattern, as shown in Fig.~\ref{fig:model}(b) for $n=3$ and $m=4$. Removing the last dangling triangle, we have a total of $2nm-1=23$ triangles. Subsequently, we systematically apply the $z$-shift monotonically at the contact hinge of every consecutive pair along the assumed sequence, resulting in a total depth of $2d(nm-1)$. Note that, if we reconfigure the lattice towards the compact state by applying a global soft mode, the signs of the twist angles alternate along the sequence, which therefore can be seen as the lattice \textit{deployment sequence}.

\begin{figure}[t!]
\includegraphics[width=\columnwidth]{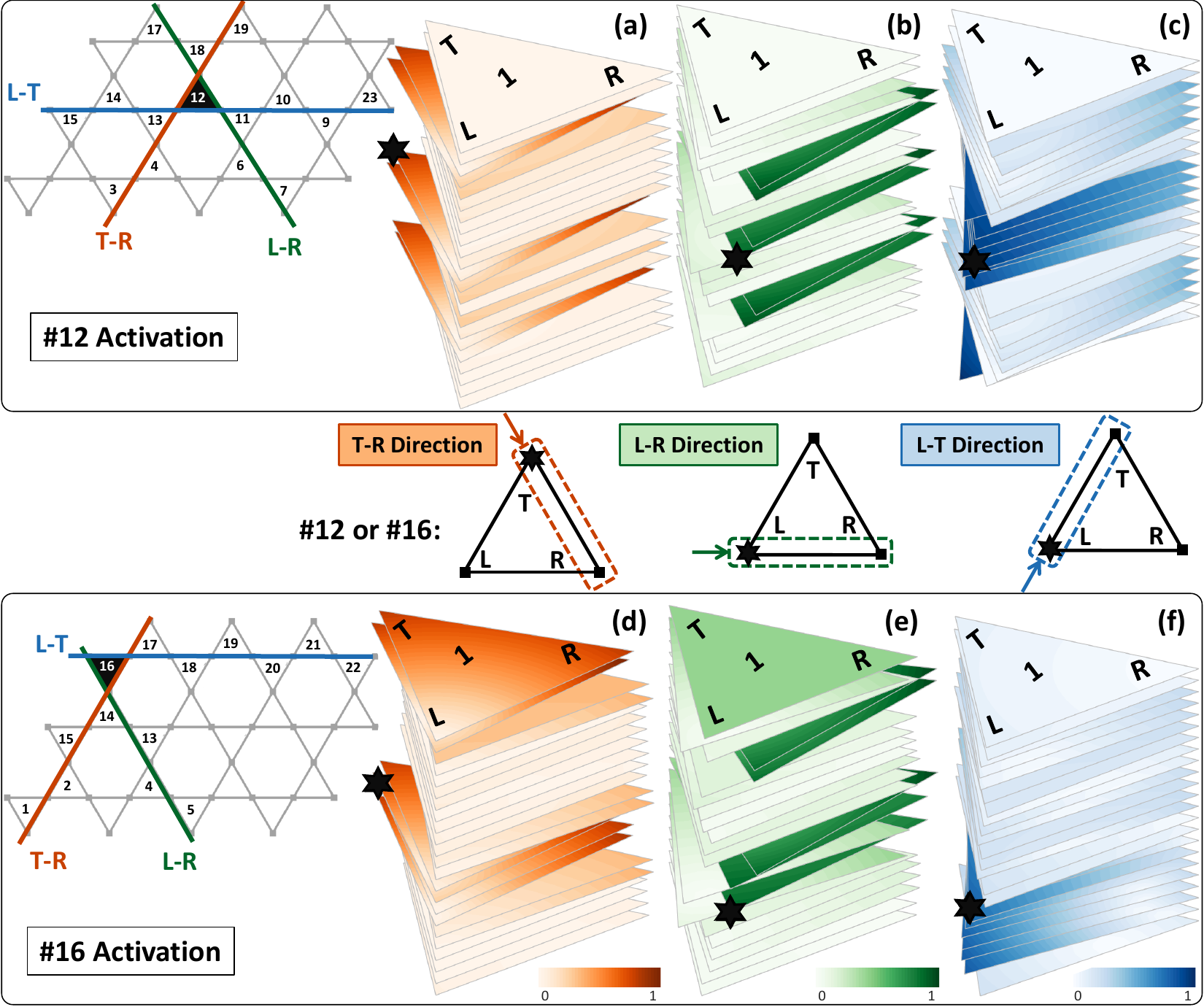}
\caption{\label{fig:simulation}Full-scale static simulation results for 23-cell chains (and corresponding deployed lattices) loaded at 
triangles $\#12$ and $\#16$ along the (a, d) T-R, (b, e) L-R, and (c, f) L-T directions. The triangles are colored according to the magnitude (normalized by highest) of their twist. 
The activated triangles match 
the selected directions in the deployed lattices.}
\end{figure}

\begin{figure*}[t]
\includegraphics[width=\textwidth]{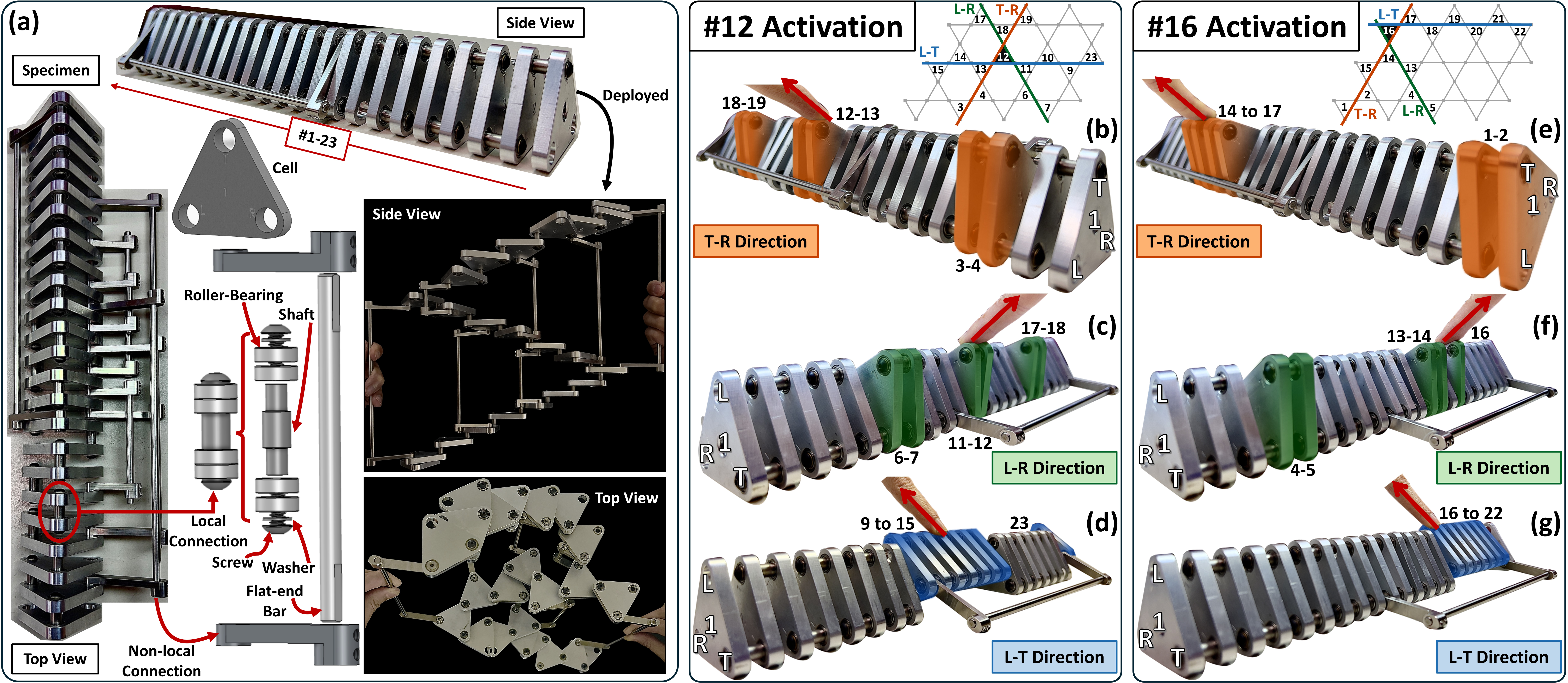}
\caption{(a) Top and angled views of the chain prototype, along with zoomed-in details of its components, highlight the technology involved in realizing non-local, long-range inter-cell connections. Insets show side and top views of the prototype during deployment. (b-g) Snapshots from static experiments with loading applied at triangle $\#12$ (left panel) and $\#16$ (right panel), along the (b,e) T-R, (c,f) L-R, and (d,g) L-T directions. Results show activation sequences of triangles corresponding to the three floppy mode directions of the deployed state. The system allows switching between local activation (d and g) and non-local activation (b-c and e-f) through a simple flip of the loading direction.}
\label{fig:exp}
\end{figure*}

In the compact case at $\theta=0^\circ$, all triangles collapse and align to form a 1D prismatic structure with a layout and kinematics akin to those of a chain capable of deforming through the relative rotation of its triangular cross-sections, shown in Fig.\ref{fig:model}(c). 
Unlike conventional torsional chains, rotations here do not occur about a shared axis, 
but rather about three distinct parallel axes passing through the hinges labeled T, L, R in Figs.~\ref{fig:model}(b-c), at the top, left, and right corners, respectively. 
One of the most striking consequences of collapsing the kagome into a chain is the emergence of \textit{non-local} interactions. According to the deployment sequence, in-plane triangles that are mechanically connected and belong to the same row of the deployed lattice occupy adjacent positions in the chain, establishing local interactions along its axis. In contrast, those belonging to different rows are positioned farther apart in the chain, resulting in non-local interactions. Alternatively, the numbering strategy dictates that some connections between adjacent pairs of triangles in the planar case involve consecutive indices and incremental $z$ coordinates in the chain, while others feature discrete jumps, marked by finite gaps in $z$. Specifically, seven non-local connections are identified and color-coded in Figs.~\ref{fig:model}(b-c). It is important to note that the physical nature of the connection does not change between Fig.~\ref{fig:model}(b) and Fig.~\ref{fig:model}(c). Nevertheless, non-locality endows selected connections with a new functionality when we interpret their kinematics from the perspective of the twist mechanics of the chain. In other words, without introducing any change in the kinematics, nearest neighbors in the 2D lattice become non-nearest neighbors along the axis of the chain (SM.\ref{sec.2}).

Duality guarantees that the phonon spectrum of the co-planar dynamics of the chain ($\theta = 0^\circ$) matches that of the in-plane dynamics of its dual counterpart ($\theta = 180^\circ$), and should therefore also exhibit bulk floppy modes. Notably, within the finite domain in Fig.~\ref{fig:model}(b), we identify 13 straight lines aligned along the T-R, L-R, and L-T directions, associated with the $\Gamma-M$ direction, that support bulk floppy modes. 
To understand how these modes manifest within the 23-cell chain, we conduct static simulations, constraining one edge of the first and last triangles and loading a randomly selected triangle ($\#12$) with a unit-amplitude force. We conduct three sets of simulations, loading the T point along the T-R direction, the L point along L-R, and the L point along L-T, respectively. The resulting twist sequences are shown in Figs.~\ref{fig:simulation}(a-c). We observe that, in each case, the twist sequence involves triangles corresponding to those clustered along the selected direction of aligned bonds passing through triangle $\#12$ in the deployed state. For instance, loading the T point along T-R (Fig.~\ref{fig:simulation}(a)) mainly induces the twist of the (3-4-12-13-18-19) sequence. Similarly, loading along L-R (Fig.~\ref{fig:simulation}(b)) activates the (6-7-11-12-17-18) sequence (see SM.\ref{sec.3} for the deployed state). Interestingly, as a result of non-locality, the activation sequence involves some sets of non-consecutive triangles. In contrast, loading along L-T results in a predominantly compact twist sequence, as the L-T triangles are linked by local connections in the chain. We repeat this exercise for triangle $\#16$, observing similar behavior, see Figs.~\ref{fig:simulation}(d-e).

\section{Experiments on kagome chain prototype}
Seeking experimental verification of the result in Fig.~\ref{fig:simulation} requires overcoming the non-trivial fabrication challenge of realizing a physical prototype endowed with all the non-local interactions required by the model. After several iterations, our design process lands on the prototype illustrated in Fig.~\ref{fig:exp}(a), shown along with schematics of the hinge components. The triangles are water-jet cut from an aluminum slab and connected to their neighbors via pin hinges. Each hinge comprises a shaft connected through washers and screws to roller bearings that guarantee virtually friction-less rotations and minimize any jiggling that would cause loss of co-axiality. The non-local interactions are realized through arm links that stem from the hinges they are intended to connect, free to rotate with respect to them but rigidly attached to long bars that are properly offset from the chain. The flat sections are tightly fastened at each end of the bar using small screws and thread lockers to prevent loosening over loading cycles. This setup allows maintaining co-axiality between the vertices that should remain connected at all stages of deployment, without interfering with those that are aligned only at $\theta=0^\circ$. Two snapshots capturing different views of the deployment of the chain are provided in the inset of Fig.~\ref{fig:exp}(a) (details in SM.\ref{sec.4}). 

\begin{figure}[t]
\includegraphics[width=\columnwidth]{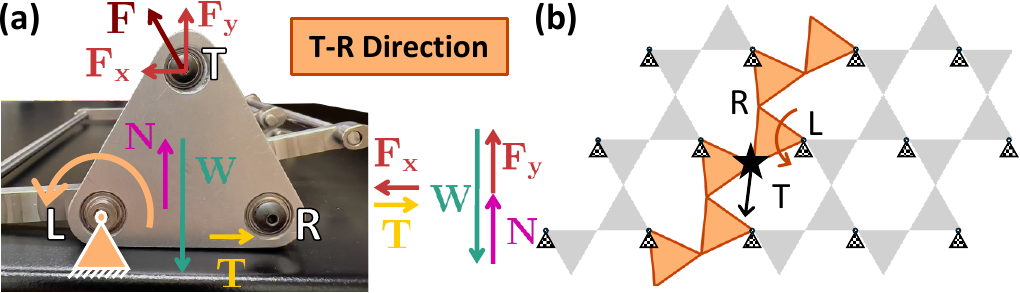}
\caption{\label{fig:NEWBC}(a) Graphical illustration of the static equilibrium considerations involved in modeling the interaction of the chain with its underlying support. The free-body diagram depicts the forces on the prototype when loading the T corner along T-R. (b) Static simulation of the corresponding deployed lattice for T-R loading, incorporating updated boundary conditions inspired by the kinematics illustrated in (a).}
\end{figure}

To subject the prototype to mechanical testing, we need to circumvent a practical challenge. As fabricated, the chain is unstable, tending to automatically deploy to its expanded state if left unconstrained. This issue arises from the interplay between gravity and the geometric eccentricity of the specimen, due to the non-local connections, which concur to exert a torque on the chain that is not resisted by the nearly frictionless hinges. The problem is especially pronounced if the chain is hung vertically. To curb the issue, we propose a setup where the chain rests horizontally on a flat surface. Figs.~\ref{fig:exp}(b-g) illustrate the behavior of the chain when loaded statically at triangle $\#12$ (Figs.~\ref{fig:exp}(b-d)) and $\#16$ (Figs.~\ref{fig:exp}(e-g)). The load is applied manually by pulling the desired corner (T or L) of the specified triangle ($\#12$ or $\#16$) along the intended direction (T-R, L-R, or L-T). Note that we orient the entire specimen according to the corner intended for loading. For instance, to load the T corner, (Figs.~\ref{fig:exp}(b, e)), the specimen is aligned such that the L-R side rests on the surface. Conversely, to load the L corner (Figs.~\ref{fig:exp}(c, d, f, g)), the T-R side is laid on the surface. 

Throughout the loading process, we ensure that the specimen is not lifted from the surface nor slides on it. This is achieved by keeping the vertical component ($\vb{F_Y}$) of the load ($\vb{F}$) not exceeding the weight of the chain ($\vb{W}$) and maintaining the horizontal reaction force ($\vb{T}$), which balances the horizontal component of the load ($\vb{F_X}$), below the maximum allowable force without sliding ($\vb{T<T_{max}}=\mu_s \vb{N}$, where $\vb{N}$ is the vertical reaction form the surface and $\mu_s$ is the coefficient of static friction). These conditions, graphically depicted in Fig.~\ref{fig:NEWBC}(a), are easily met working with a sufficiently frictional surface and a heavy chain (as in our study). Under these equilibrium conditions, the load boils down to a moment about one of the vertices resting on the surface, which in the absence of sliding, can be treated as a pin support (details of SolidWorks animations in SM.\ref{sec.5}). Accounting for these kinematics of the setup requires a revision of the boundary conditions with respect to those assumed in the simulations of Fig.~\ref{fig:simulation} (SM.\ref{sec.6}). This adjustment can be enforced pinning all the triangle corners of the lattice that, in the deployed state, do not lie along lattice directions parallel to the applied load. For example, all L corners should be pinned when we load along the T-R direction. A simulation of the deployed state, incorporating the revised boundary conditions, is shown in Fig.~\ref{fig:NEWBC}(b) for T-R loading, showing the manifestation of soft modes under the additional constraints.  For simulations for other loading cases see SM.\ref{sec.7}.

\begin{figure}[t!]
\includegraphics[width=\columnwidth]{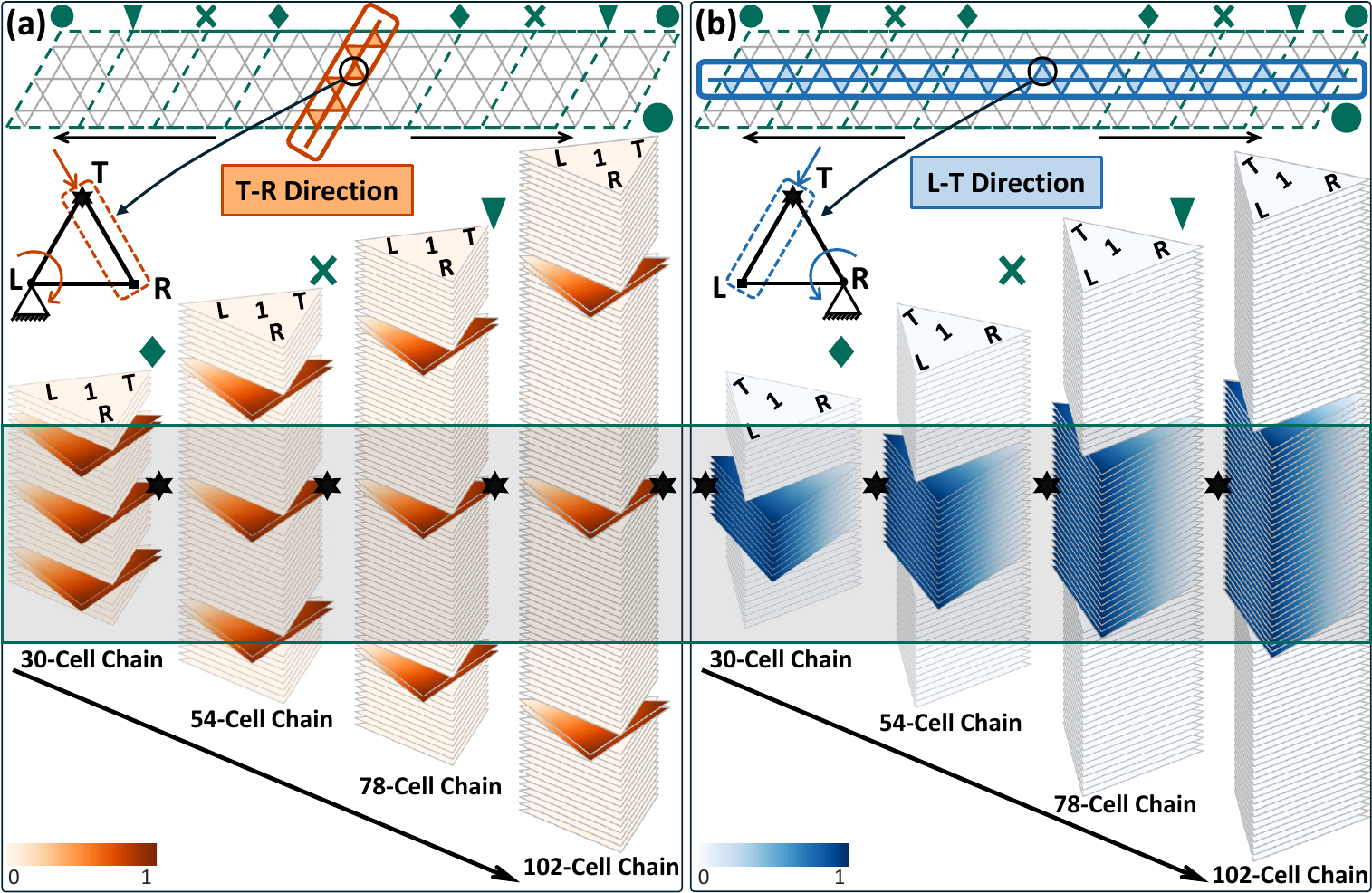}
\caption{\label{fig:sparsity}(a-b) Parametric study revealing the exacerbation of the dichotomy between local and non-local twist sequences in chains of increasing cell numbers. The deployed configurations of the different chains depicted at the top of each panel show that longer chains correspond to finite lattices with more columns and equal numbers of rows. The shaded box outlines a 31-cell interval around the load. As the chain lengthens, the non-local activation sequences shift progressively away from the load, while the local ones keep clustering.}
\end{figure}

With the kinematics of the horizontal chain figured out, we can interpret the results shown in Figs.~\ref{fig:exp}(b-d) for triangle $\#12$. We observe that the behavior agrees with the predictions from band diagram and static simulations. For each direction of loading, only the triangles that are located along the loading direction in the deployed state are activated along the chain. This effect has an unexpected consequence. Since the modes directed along the three floppy directions involve different sets of triangles, which are differently distributed along the length of the chain, the selection of a direction of loading results in the activation of qualitatively different twist sequences. Those along T-R and L-R involve distributed, non-local twist patterns, while the one along L-T results by and large in a localized, clustered pattern. In essence, the chain displays an ability to switch between local and non-local activation sequences that is controlled solely by the selected direction of loading. Similar considerations can be made for triangle $\#16$ in Figs.~\ref{fig:exp}(e-g). 

\section{Discussion on load direction selectivity}
In order to emphasize the ability of the system to switch between local and non-local effects based solely on the direction of loading, let us revisit the problem using longer chains associated with larger lattice domains. We simulate 
chains with increasing cell numbers (30, 54, 78, 102), as shown in Fig.~\ref{fig:sparsity}. We start with a 30-cell chain, corresponding to a $3 \times 5$ deployed configuration. For each subsequent scenario, we add four columns to the previous deployed configuration and we determine the necessary hinge connectivity for the corresponding chain. We then subject the center triangle of each chain to a unit-amplitude force applied at the T corner, targeting the T-R and L-T loading directions. The resulting twist patterns for T-R and L-T are depicted in Fig.~\ref{fig:sparsity}(a) and Fig.~\ref{fig:sparsity}(b), respectively. Observation of the results allows us to formulate a simple rule governing the relation between the activation sequence and the lattice dimensionality and layout. For a finite $n-$row by $m-$column domain, T-R or L-R loading yields a distributed response featuring $n$ peaks, each encompassing two triangles, while L-T loading activates a localized response involving a cluster of $2m$ triangles.

We noted above that varying the loading direction can lead to an abrupt switch between localized and distributed activation sequences. As the chain length increases, non-local activations due to T-R loading propagate further from the loaded cell. This is quantified by considering a control window encompassing approximately 31 cells, centered around the loaded cell, as shown in Fig.~\ref{fig:sparsity}. In contrast, for L-T loading, the twisted cluster remains localized and tends to occupy a progressively larger portion of the control window. In other words, the dichotomy between local and non-local effects grows. In the limit of an infinite lattice, when we load along T-R, all the non-local activations shift indefinitely outwards, away from the loading spot, and the response reduces to the twist of the loaded triangle and the adjacent one. When we load along L-T, the cluster of twisted triangles grows indefinitely, encompassing an increasingly long portion of the chain about the loading point. In summary, we establish a total dichotomy between an extremely localized response for T-R loading and a distributed response for L-T loading. Such dichotomy can only be appreciated through the lens of the chain dynamics, where the interaction landscape embedded in the lattice cell sequence manifests in the form of non-locality.

\section{Conclusion}
In conclusion, in this work we documented the emergence of a series of unintuitive activations of floppy mechanisms resulting from emerging non-locality in kagome chains obtained by collapsing regular kagome lattices to their fully compact states via the Guest-Hutchinson mode. The non-locality produces marked dichotomies in the chain mechanisms that are sensitive to the direction of loading, yielding an unprecedented selective character of the chain torsional response.

\section{acknowledgement}
The authors acknowledge support from the National Science Foundation (grant CMMI-2027000). They are grateful to K. Sun and S. Sarkar (U. of Michigan) for insightful discussions, and to Peter Ness and Max Olson (U. of Minnesota CSE Shop) for precious assistance with prototype design and fabrication.


\bibliographystyle{apsrev4-1}
\bibliography{ref}

\begin{thebibliography}{35}%
\makeatletter
\providecommand \@ifxundefined [1]{%
 \@ifx{#1\undefined}
}%
\providecommand \@ifnum [1]{%
 \ifnum #1\expandafter \@firstoftwo
 \else \expandafter \@secondoftwo
 \fi
}%
\providecommand \@ifx [1]{%
 \ifx #1\expandafter \@firstoftwo
 \else \expandafter \@secondoftwo
 \fi
}%
\providecommand \natexlab [1]{#1}%
\providecommand \enquote  [1]{``#1''}%
\providecommand \bibnamefont  [1]{#1}%
\providecommand \bibfnamefont [1]{#1}%
\providecommand \citenamefont [1]{#1}%
\providecommand \href@noop [0]{\@secondoftwo}%
\providecommand \href [0]{\begingroup \@sanitize@url \@href}%
\providecommand \@href[1]{\@@startlink{#1}\@@href}%
\providecommand \@@href[1]{\endgroup#1\@@endlink}%
\providecommand \@sanitize@url [0]{\catcode `\\12\catcode `\$12\catcode `\&12\catcode `\#12\catcode `\^12\catcode `\_12\catcode `\%12\relax}%
\providecommand \@@startlink[1]{}%
\providecommand \@@endlink[0]{}%
\providecommand \url  [0]{\begingroup\@sanitize@url \@url }%
\providecommand \@url [1]{\endgroup\@href {#1}{\urlprefix }}%
\providecommand \urlprefix  [0]{URL }%
\providecommand \Eprint [0]{\href }%
\providecommand \doibase [0]{http://dx.doi.org/}%
\providecommand \selectlanguage [0]{\@gobble}%
\providecommand \bibinfo  [0]{\@secondoftwo}%
\providecommand \bibfield  [0]{\@secondoftwo}%
\providecommand \translation [1]{[#1]}%
\providecommand \BibitemOpen [0]{}%
\providecommand \bibitemStop [0]{}%
\providecommand \bibitemNoStop [0]{.\EOS\space}%
\providecommand \EOS [0]{\spacefactor3000\relax}%
\providecommand \BibitemShut  [1]{\csname bibitem#1\endcsname}%
\let\auto@bib@innerbib\@empty
\bibitem [{\citenamefont {Lee}\ \emph {et~al.}(2012)\citenamefont {Lee}, \citenamefont {Singer},\ and\ \citenamefont {Thomas}}]{Lee_et_all_mech_mat_AdvMat_2012}%
  \BibitemOpen
  \bibfield  {author} {\bibinfo {author} {\bibfnamefont {J.-H.}\ \bibnamefont {Lee}}, \bibinfo {author} {\bibfnamefont {J.~P.}\ \bibnamefont {Singer}}, \ and\ \bibinfo {author} {\bibfnamefont {E.~L.}\ \bibnamefont {Thomas}},\ }\href {\doibase https://doi.org/10.1002/adma.201201644} {\bibfield  {journal} {\bibinfo  {journal} {Advanced Materials}\ }\textbf {\bibinfo {volume} {24}},\ \bibinfo {pages} {4782} (\bibinfo {year} {2012})}\BibitemShut {NoStop}%
\bibitem [{\citenamefont {Babaee}\ \emph {et~al.}(2013)\citenamefont {Babaee}, \citenamefont {Shim}, \citenamefont {Weaver}, \citenamefont {Chen}, \citenamefont {Patel},\ and\ \citenamefont {Bertoldi}}]{Babaee_et_sll_metamaterial_2013}%
  \BibitemOpen
  \bibfield  {author} {\bibinfo {author} {\bibfnamefont {S.}~\bibnamefont {Babaee}}, \bibinfo {author} {\bibfnamefont {J.}~\bibnamefont {Shim}}, \bibinfo {author} {\bibfnamefont {J.~C.}\ \bibnamefont {Weaver}}, \bibinfo {author} {\bibfnamefont {E.~R.}\ \bibnamefont {Chen}}, \bibinfo {author} {\bibfnamefont {N.}~\bibnamefont {Patel}}, \ and\ \bibinfo {author} {\bibfnamefont {K.}~\bibnamefont {Bertoldi}},\ }\href {\doibase https://doi.org/10.1002/adma.201301986} {\bibfield  {journal} {\bibinfo  {journal} {Advanced Materials}\ }\textbf {\bibinfo {volume} {25}},\ \bibinfo {pages} {5044} (\bibinfo {year} {2013})}\BibitemShut {NoStop}%
\bibitem [{\citenamefont {Christensen}\ \emph {et~al.}(2015)\citenamefont {Christensen}, \citenamefont {Kadic}, \citenamefont {Wegener}, \citenamefont {Kraft},\ and\ \citenamefont {Wegener}}]{Christensen_et_ell_MechMat_2015}%
  \BibitemOpen
  \bibfield  {author} {\bibinfo {author} {\bibfnamefont {J.}~\bibnamefont {Christensen}}, \bibinfo {author} {\bibfnamefont {M.}~\bibnamefont {Kadic}}, \bibinfo {author} {\bibfnamefont {M.}~\bibnamefont {Wegener}}, \bibinfo {author} {\bibfnamefont {O.}~\bibnamefont {Kraft}}, \ and\ \bibinfo {author} {\bibfnamefont {M.}~\bibnamefont {Wegener}},\ }\href {\doibase 10.1557/mrc.2015.51} {\bibfield  {journal} {\bibinfo  {journal} {MRS Communications}\ }\textbf {\bibinfo {volume} {5}},\ \bibinfo {pages} {453} (\bibinfo {year} {2015})}\BibitemShut {NoStop}%
\bibitem [{\citenamefont {Rafsanjani}\ and\ \citenamefont {Pasini}(2016)}]{Rafsanjani_Pasini__metamaterials_2016}%
  \BibitemOpen
  \bibfield  {author} {\bibinfo {author} {\bibfnamefont {A.}~\bibnamefont {Rafsanjani}}\ and\ \bibinfo {author} {\bibfnamefont {D.}~\bibnamefont {Pasini}},\ }\href {http://dx.doi.org/10.1016/j.eml.2016.09.001} {\bibfield  {journal} {\bibinfo  {journal} {Extreme Mechanics Letters}\ }\textbf {\bibinfo {volume} {9}},\ \bibinfo {pages} {291} (\bibinfo {year} {2016})}\BibitemShut {NoStop}%
\bibitem [{\citenamefont {Bertoldi}\ \emph {et~al.}(2017{\natexlab{a}})\citenamefont {Bertoldi}, \citenamefont {Vitelli}, \citenamefont {Christensen},\ and\ \citenamefont {van Hecke}}]{metamaterials_Bertoldi2017}%
  \BibitemOpen
  \bibfield  {author} {\bibinfo {author} {\bibfnamefont {K.}~\bibnamefont {Bertoldi}}, \bibinfo {author} {\bibfnamefont {V.}~\bibnamefont {Vitelli}}, \bibinfo {author} {\bibfnamefont {J.}~\bibnamefont {Christensen}}, \ and\ \bibinfo {author} {\bibfnamefont {M.}~\bibnamefont {van Hecke}},\ }\href {\doibase 10.1038/natrevmats.2017.66} {\bibfield  {journal} {\bibinfo  {journal} {Nature Reviews Materials}\ }\textbf {\bibinfo {volume} {2}} (\bibinfo {year} {2017}{\natexlab{a}}),\ 10.1038/natrevmats.2017.66}\BibitemShut {NoStop}%
\bibitem [{\citenamefont {Deng}\ \emph {et~al.}(2017)\citenamefont {Deng}, \citenamefont {Raney}, \citenamefont {Tournat},\ and\ \citenamefont {Bertoldi}}]{Deng_et_all_soliton_PRL2017}%
  \BibitemOpen
  \bibfield  {author} {\bibinfo {author} {\bibfnamefont {B.}~\bibnamefont {Deng}}, \bibinfo {author} {\bibfnamefont {J.~R.}\ \bibnamefont {Raney}}, \bibinfo {author} {\bibfnamefont {V.}~\bibnamefont {Tournat}}, \ and\ \bibinfo {author} {\bibfnamefont {K.}~\bibnamefont {Bertoldi}},\ }\href {\doibase 10.1103/PhysRevLett.118.204102} {\bibfield  {journal} {\bibinfo  {journal} {Phys. Rev. Lett.}\ }\textbf {\bibinfo {volume} {118}},\ \bibinfo {pages} {204102} (\bibinfo {year} {2017})}\BibitemShut {NoStop}%
\bibitem [{\citenamefont {Deng}\ \emph {et~al.}(2020)\citenamefont {Deng}, \citenamefont {Yu}, \citenamefont {Forte}, \citenamefont {Tournat},\ and\ \citenamefont {Bertoldi}}]{Bolei_et_all_floppy_PNAS2020}%
  \BibitemOpen
  \bibfield  {author} {\bibinfo {author} {\bibfnamefont {B.}~\bibnamefont {Deng}}, \bibinfo {author} {\bibfnamefont {S.}~\bibnamefont {Yu}}, \bibinfo {author} {\bibfnamefont {A.~E.}\ \bibnamefont {Forte}}, \bibinfo {author} {\bibfnamefont {V.}~\bibnamefont {Tournat}}, \ and\ \bibinfo {author} {\bibfnamefont {K.}~\bibnamefont {Bertoldi}},\ }\href {\doibase 10.1073/pnas.2015847117} {\bibfield  {journal} {\bibinfo  {journal} {Proceedings of the National Academy of Sciences}\ }\textbf {\bibinfo {volume} {117}},\ \bibinfo {pages} {31002} (\bibinfo {year} {2020})},\ \Eprint {http://arxiv.org/abs/https://www.pnas.org/doi/pdf/10.1073/pnas.2015847117} {https://www.pnas.org/doi/pdf/10.1073/pnas.2015847117} \BibitemShut {NoStop}%
\bibitem [{\citenamefont {Brillouin}(1953)}]{brillouin_wave_1953}%
  \BibitemOpen
  \bibfield  {author} {\bibinfo {author} {\bibfnamefont {L.}~\bibnamefont {Brillouin}},\ }\href {https://books.google.com/books?id=jOQNAQAAIAAJ} {\emph {\bibinfo {title} {Wave Propagation in Periodic Structures: Electric Filters and Crystal Lattices}}}\ (\bibinfo  {publisher} {Dover Publications},\ \bibinfo {year} {1953})\BibitemShut {NoStop}%
\bibitem [{\citenamefont {Phani}\ \emph {et~al.}(2006)\citenamefont {Phani}, \citenamefont {Woodhouse},\ and\ \citenamefont {Fleck}}]{Phani_et_all_Acoustical_Soc_America_2006}%
  \BibitemOpen
  \bibfield  {author} {\bibinfo {author} {\bibfnamefont {A.~S.}\ \bibnamefont {Phani}}, \bibinfo {author} {\bibfnamefont {J.}~\bibnamefont {Woodhouse}}, \ and\ \bibinfo {author} {\bibfnamefont {N.~A.}\ \bibnamefont {Fleck}},\ }\href {\doibase 10.1121/1.2179748} {\bibfield  {journal} {\bibinfo  {journal} {The Journal of the Acoustical Society of America}\ }\textbf {\bibinfo {volume} {119}},\ \bibinfo {pages} {1995} (\bibinfo {year} {2006})}\BibitemShut {NoStop}%
\bibitem [{\citenamefont {Czajkowski}\ \emph {et~al.}(2022)\citenamefont {Czajkowski}, \citenamefont {Coulais}, \citenamefont {van Hecke},\ and\ \citenamefont {Rocklin}}]{mechanism_metamaterials_Coulais_Natcom2022}%
  \BibitemOpen
  \bibfield  {author} {\bibinfo {author} {\bibfnamefont {M.}~\bibnamefont {Czajkowski}}, \bibinfo {author} {\bibfnamefont {C.}~\bibnamefont {Coulais}}, \bibinfo {author} {\bibfnamefont {M.}~\bibnamefont {van Hecke}}, \ and\ \bibinfo {author} {\bibfnamefont {D.~Z.}\ \bibnamefont {Rocklin}},\ }\href {\doibase 10.1038/s41467-021-27825-0} {\bibfield  {journal} {\bibinfo  {journal} {Nature Communications}\ }\textbf {\bibinfo {volume} {13}} (\bibinfo {year} {2022}),\ 10.1038/s41467-021-27825-0}\BibitemShut {NoStop}%
\bibitem [{\citenamefont {Maxwell}(1864)}]{Maxwell_1864}%
  \BibitemOpen
  \bibfield  {author} {\bibinfo {author} {\bibfnamefont {J.~C.}\ \bibnamefont {Maxwell}},\ }\href {\doibase 10.1080/14786446408643668} {\bibfield  {journal} {\bibinfo  {journal} {The London, Edinburgh, and Dublin Philosophical Magazine and Journal of Science}\ }\textbf {\bibinfo {volume} {27}},\ \bibinfo {pages} {294} (\bibinfo {year} {1864})}\BibitemShut {NoStop}%
\bibitem [{\citenamefont {Calladine}(1978)}]{calladine_lattice_IJSS_1978}%
  \BibitemOpen
  \bibfield  {author} {\bibinfo {author} {\bibfnamefont {C.}~\bibnamefont {Calladine}},\ }\href {\doibase 10.1016/0020-7683(78)90052-5} {\bibfield  {journal} {\bibinfo  {journal} {International Journal of Solids and Structures}\ }\textbf {\bibinfo {volume} {14}},\ \bibinfo {pages} {161} (\bibinfo {year} {1978})}\BibitemShut {NoStop}%
\bibitem [{\citenamefont {Thorpe}(1983)}]{Thorpe_floppy_1983}%
  \BibitemOpen
  \bibfield  {author} {\bibinfo {author} {\bibfnamefont {M.~F.}\ \bibnamefont {Thorpe}},\ }\href {\doibase 10.1016/0022-3093(83)90424-6} {\bibfield  {journal} {\bibinfo  {journal} {Journal of Non-Crystalline Solids}\ }\textbf {\bibinfo {volume} {57}},\ \bibinfo {pages} {355} (\bibinfo {year} {1983})}\BibitemShut {NoStop}%
\bibitem [{\citenamefont {Connelly}(2005)}]{Connelly2005}%
  \BibitemOpen
  \bibfield  {author} {\bibinfo {author} {\bibfnamefont {R.}~\bibnamefont {Connelly}},\ }\href {\doibase 10.1007/s00454-004-1124-4} {\bibfield  {journal} {\bibinfo  {journal} {Discrete \& Computational Geometry}\ }\textbf {\bibinfo {volume} {33}},\ \bibinfo {pages} {549} (\bibinfo {year} {2005})}\BibitemShut {NoStop}%
\bibitem [{\citenamefont {Lubensky}\ \emph {et~al.}(2015)\citenamefont {Lubensky}, \citenamefont {Kane}, \citenamefont {Mao}, \citenamefont {Souslov},\ and\ \citenamefont {Sun}}]{Lubensky_et_all_2015}%
  \BibitemOpen
  \bibfield  {author} {\bibinfo {author} {\bibfnamefont {T.~C.}\ \bibnamefont {Lubensky}}, \bibinfo {author} {\bibfnamefont {C.~L.}\ \bibnamefont {Kane}}, \bibinfo {author} {\bibfnamefont {X.}~\bibnamefont {Mao}}, \bibinfo {author} {\bibfnamefont {A.}~\bibnamefont {Souslov}}, \ and\ \bibinfo {author} {\bibfnamefont {K.}~\bibnamefont {Sun}},\ }\href {\doibase 10.1088/0034-4885/78/7/073901} {\bibfield  {journal} {\bibinfo  {journal} {Reports on Progress in Physics}\ }\textbf {\bibinfo {volume} {78}},\ \bibinfo {pages} {073901} (\bibinfo {year} {2015})}\BibitemShut {NoStop}%
\bibitem [{\citenamefont {Mao}\ and\ \citenamefont {Lubensky}(2011)}]{Mao_Lubensky_APS_2011}%
  \BibitemOpen
  \bibfield  {author} {\bibinfo {author} {\bibfnamefont {X.}~\bibnamefont {Mao}}\ and\ \bibinfo {author} {\bibfnamefont {T.~C.}\ \bibnamefont {Lubensky}},\ }\href {\doibase 10.1103/PhysRevE.83.011111} {\bibfield  {journal} {\bibinfo  {journal} {Phys. Rev. E}\ }\textbf {\bibinfo {volume} {83}},\ \bibinfo {pages} {011111} (\bibinfo {year} {2011})}\BibitemShut {NoStop}%
\bibitem [{\citenamefont {Kane}\ and\ \citenamefont {Lubensky}(2014)}]{Kane_Lubensky_Nphys_2014}%
  \BibitemOpen
  \bibfield  {author} {\bibinfo {author} {\bibfnamefont {C.~L.}\ \bibnamefont {Kane}}\ and\ \bibinfo {author} {\bibfnamefont {T.~C.}\ \bibnamefont {Lubensky}},\ }\href {\doibase 10.1038/nphys2835} {\bibfield  {journal} {\bibinfo  {journal} {Nature Physics}\ }\textbf {\bibinfo {volume} {10}},\ \bibinfo {pages} {39} (\bibinfo {year} {2014})}\BibitemShut {NoStop}%
\bibitem [{\citenamefont {Mao}\ and\ \citenamefont {Lubensky}(2018)}]{Mao_Lubensky_maxwell_topo_2018}%
  \BibitemOpen
  \bibfield  {author} {\bibinfo {author} {\bibfnamefont {X.}~\bibnamefont {Mao}}\ and\ \bibinfo {author} {\bibfnamefont {T.~C.}\ \bibnamefont {Lubensky}},\ }\href {\doibase 10.1146/annurev-conmatphys-033117-054235} {\bibfield  {journal} {\bibinfo  {journal} {Annual Review of Condensed Matter Physics}\ }\textbf {\bibinfo {volume} {9}},\ \bibinfo {pages} {413} (\bibinfo {year} {2018})}\BibitemShut {NoStop}%
\bibitem [{\citenamefont {Schaeffer}\ and\ \citenamefont {Ruzzene}(2015)}]{Schaeffer_Ruzzene_kagome_appliedphys_2015}%
  \BibitemOpen
  \bibfield  {author} {\bibinfo {author} {\bibfnamefont {M.}~\bibnamefont {Schaeffer}}\ and\ \bibinfo {author} {\bibfnamefont {M.}~\bibnamefont {Ruzzene}},\ }\href {\doibase 10.1063/1.4921358} {\bibfield  {journal} {\bibinfo  {journal} {Journal of Applied Physics}\ }\textbf {\bibinfo {volume} {117}},\ \bibinfo {pages} {194903} (\bibinfo {year} {2015})}\BibitemShut {NoStop}%
\bibitem [{\citenamefont {Bertoldi}\ \emph {et~al.}(2017{\natexlab{b}})\citenamefont {Bertoldi}, \citenamefont {Vitelli}, \citenamefont {Christensen},\ and\ \citenamefont {van Hecke}}]{Bertoldi_et_all_natrevmats_2017}%
  \BibitemOpen
  \bibfield  {author} {\bibinfo {author} {\bibfnamefont {K.}~\bibnamefont {Bertoldi}}, \bibinfo {author} {\bibfnamefont {V.}~\bibnamefont {Vitelli}}, \bibinfo {author} {\bibfnamefont {J.}~\bibnamefont {Christensen}}, \ and\ \bibinfo {author} {\bibfnamefont {M.}~\bibnamefont {van Hecke}},\ }\href {\doibase 10.1038/natrevmats.2017.66} {\bibfield  {journal} {\bibinfo  {journal} {Nature Reviews Materials}\ }\textbf {\bibinfo {volume} {2}},\ \bibinfo {pages} {17066} (\bibinfo {year} {2017}{\natexlab{b}})}\BibitemShut {NoStop}%
\bibitem [{\citenamefont {Chen}\ \emph {et~al.}(2018)\citenamefont {Chen}, \citenamefont {Nassar}, \citenamefont {Norris}, \citenamefont {Hu},\ and\ \citenamefont {Huang}}]{Chen_et_all_kagome_PhysRevB_2018}%
  \BibitemOpen
  \bibfield  {author} {\bibinfo {author} {\bibfnamefont {H.}~\bibnamefont {Chen}}, \bibinfo {author} {\bibfnamefont {H.}~\bibnamefont {Nassar}}, \bibinfo {author} {\bibfnamefont {A.~N.}\ \bibnamefont {Norris}}, \bibinfo {author} {\bibfnamefont {G.~K.}\ \bibnamefont {Hu}}, \ and\ \bibinfo {author} {\bibfnamefont {G.~L.}\ \bibnamefont {Huang}},\ }\href {\doibase 10.1103/PhysRevB.98.094302} {\bibfield  {journal} {\bibinfo  {journal} {Phys. Rev. B}\ }\textbf {\bibinfo {volume} {98}},\ \bibinfo {pages} {094302} (\bibinfo {year} {2018})}\BibitemShut {NoStop}%
\bibitem [{\citenamefont {Riva}\ \emph {et~al.}(2018)\citenamefont {Riva}, \citenamefont {Quadrelli}, \citenamefont {Cazzulani},\ and\ \citenamefont {Braghin}}]{Riva_et_all_topo_kagome_appliedphys_2018}%
  \BibitemOpen
  \bibfield  {author} {\bibinfo {author} {\bibfnamefont {E.}~\bibnamefont {Riva}}, \bibinfo {author} {\bibfnamefont {D.~E.}\ \bibnamefont {Quadrelli}}, \bibinfo {author} {\bibfnamefont {G.}~\bibnamefont {Cazzulani}}, \ and\ \bibinfo {author} {\bibfnamefont {F.}~\bibnamefont {Braghin}},\ }\href {\doibase 10.1063/1.5045837} {\bibfield  {journal} {\bibinfo  {journal} {Journal of Applied Physics}\ }\textbf {\bibinfo {volume} {124}},\ \bibinfo {pages} {164903} (\bibinfo {year} {2018})}\BibitemShut {NoStop}%
\bibitem [{\citenamefont {Nassar}\ \emph {et~al.}(2020)\citenamefont {Nassar}, \citenamefont {Chen},\ and\ \citenamefont {Huang}}]{Nassar-et-al_Microtwist_JMPS_2020}%
  \BibitemOpen
  \bibfield  {author} {\bibinfo {author} {\bibfnamefont {H.}~\bibnamefont {Nassar}}, \bibinfo {author} {\bibfnamefont {H.}~\bibnamefont {Chen}}, \ and\ \bibinfo {author} {\bibfnamefont {G.}~\bibnamefont {Huang}},\ }\href {\doibase https://doi.org/10.1016/j.jmps.2020.104107} {\bibfield  {journal} {\bibinfo  {journal} {Journal of the Mechanics and Physics of Solids}\ }\textbf {\bibinfo {volume} {144}},\ \bibinfo {pages} {104107} (\bibinfo {year} {2020})}\BibitemShut {NoStop}%
\bibitem [{\citenamefont {Souslov}\ \emph {et~al.}(2009)\citenamefont {Souslov}, \citenamefont {Liu},\ and\ \citenamefont {Lubensky}}]{Anton_et_all_kagome_prl_2009}%
  \BibitemOpen
  \bibfield  {author} {\bibinfo {author} {\bibfnamefont {A.}~\bibnamefont {Souslov}}, \bibinfo {author} {\bibfnamefont {A.~J.}\ \bibnamefont {Liu}}, \ and\ \bibinfo {author} {\bibfnamefont {T.~C.}\ \bibnamefont {Lubensky}},\ }\href {\doibase 10.1103/PhysRevLett.103.205503} {\bibfield  {journal} {\bibinfo  {journal} {Phys. Rev. Lett.}\ }\textbf {\bibinfo {volume} {103}},\ \bibinfo {pages} {205503} (\bibinfo {year} {2009})}\BibitemShut {NoStop}%
\bibitem [{\citenamefont {Sun}\ \emph {et~al.}(2012)\citenamefont {Sun}, \citenamefont {Souslov}, \citenamefont {Mao},\ and\ \citenamefont {Lubensky}}]{Sun_et_all_pnas_2012}%
  \BibitemOpen
  \bibfield  {author} {\bibinfo {author} {\bibfnamefont {K.}~\bibnamefont {Sun}}, \bibinfo {author} {\bibfnamefont {A.}~\bibnamefont {Souslov}}, \bibinfo {author} {\bibfnamefont {X.}~\bibnamefont {Mao}}, \ and\ \bibinfo {author} {\bibfnamefont {T.~C.}\ \bibnamefont {Lubensky}},\ }\href {\doibase 10.1073/pnas.1119941109} {\bibfield  {journal} {\bibinfo  {journal} {Proceedings of the National Academy of Sciences}\ }\textbf {\bibinfo {volume} {109}},\ \bibinfo {pages} {12369} (\bibinfo {year} {2012})}\BibitemShut {NoStop}%
\bibitem [{\citenamefont {Guest}\ and\ \citenamefont {Hutchinson}(2003)}]{GUEST_Hutchinson_2003}%
  \BibitemOpen
  \bibfield  {author} {\bibinfo {author} {\bibfnamefont {S.}~\bibnamefont {Guest}}\ and\ \bibinfo {author} {\bibfnamefont {J.}~\bibnamefont {Hutchinson}},\ }\href {\doibase https://doi.org/10.1016/S0022-5096(02)00107-2} {\bibfield  {journal} {\bibinfo  {journal} {Journal of the Mechanics and Physics of Solids}\ }\textbf {\bibinfo {volume} {51}},\ \bibinfo {pages} {383} (\bibinfo {year} {2003})}\BibitemShut {NoStop}%
\bibitem [{\citenamefont {Rocklin}\ \emph {et~al.}(2017)\citenamefont {Rocklin}, \citenamefont {Zhou}, \citenamefont {Sun},\ and\ \citenamefont {Mao}}]{Rocklin_et_all_natcom_2017}%
  \BibitemOpen
  \bibfield  {author} {\bibinfo {author} {\bibfnamefont {D.~Z.}\ \bibnamefont {Rocklin}}, \bibinfo {author} {\bibfnamefont {S.}~\bibnamefont {Zhou}}, \bibinfo {author} {\bibfnamefont {K.}~\bibnamefont {Sun}}, \ and\ \bibinfo {author} {\bibfnamefont {X.}~\bibnamefont {Mao}},\ }\href {\doibase 10.1038/ncomms14201} {\bibfield  {journal} {\bibinfo  {journal} {Nature Communications}\ }\textbf {\bibinfo {volume} {8}} (\bibinfo {year} {2017}),\ 10.1038/ncomms14201}\BibitemShut {NoStop}%
\bibitem [{\citenamefont {Li}\ and\ \citenamefont {Kohn}(2023)}]{Li_Kohn_GHkagome_JMPS2023}%
  \BibitemOpen
  \bibfield  {author} {\bibinfo {author} {\bibfnamefont {X.}~\bibnamefont {Li}}\ and\ \bibinfo {author} {\bibfnamefont {R.~V.}\ \bibnamefont {Kohn}},\ }\href {\doibase https://doi.org/10.1016/j.jmps.2023.105311} {\bibfield  {journal} {\bibinfo  {journal} {Journal of the Mechanics and Physics of Solids}\ }\textbf {\bibinfo {volume} {178}},\ \bibinfo {pages} {105311} (\bibinfo {year} {2023})}\BibitemShut {NoStop}%
\bibitem [{\citenamefont {Fruchart}\ \emph {et~al.}(2020)\citenamefont {Fruchart}, \citenamefont {Zhou},\ and\ \citenamefont {Vitelli}}]{Fruchart_et_all_duality_nat_2020}%
  \BibitemOpen
  \bibfield  {author} {\bibinfo {author} {\bibfnamefont {M.}~\bibnamefont {Fruchart}}, \bibinfo {author} {\bibfnamefont {Y.}~\bibnamefont {Zhou}}, \ and\ \bibinfo {author} {\bibfnamefont {V.}~\bibnamefont {Vitelli}},\ }\href {\doibase 10.1038/s41586-020-1932-6} {\bibfield  {journal} {\bibinfo  {journal} {Nature}\ }\textbf {\bibinfo {volume} {577}},\ \bibinfo {pages} {636} (\bibinfo {year} {2020})}\BibitemShut {NoStop}%
\bibitem [{\citenamefont {Azizi}\ \emph {et~al.}(2023)\citenamefont {Azizi}, \citenamefont {Sarkar}, \citenamefont {Sun},\ and\ \citenamefont {Gonella}}]{Azizi_et_all_PhysRevLett_2023}%
  \BibitemOpen
  \bibfield  {author} {\bibinfo {author} {\bibfnamefont {P.}~\bibnamefont {Azizi}}, \bibinfo {author} {\bibfnamefont {S.}~\bibnamefont {Sarkar}}, \bibinfo {author} {\bibfnamefont {K.}~\bibnamefont {Sun}}, \ and\ \bibinfo {author} {\bibfnamefont {S.}~\bibnamefont {Gonella}},\ }\href {\doibase 10.1103/PhysRevLett.130.156101} {\bibfield  {journal} {\bibinfo  {journal} {Phys. Rev. Lett.}\ }\textbf {\bibinfo {volume} {130}},\ \bibinfo {pages} {156101} (\bibinfo {year} {2023})}\BibitemShut {NoStop}%
\bibitem [{\citenamefont {Gonella}(2020)}]{Gonella_duality_PhysRevB_2020}%
  \BibitemOpen
  \bibfield  {author} {\bibinfo {author} {\bibfnamefont {S.}~\bibnamefont {Gonella}},\ }\href {\doibase 10.1103/PhysRevB.102.140301} {\bibfield  {journal} {\bibinfo  {journal} {Phys. Rev. B}\ }\textbf {\bibinfo {volume} {102}},\ \bibinfo {pages} {140301} (\bibinfo {year} {2020})}\BibitemShut {NoStop}%
\bibitem [{\citenamefont {Lei}\ \emph {et~al.}(2022)\citenamefont {Lei}, \citenamefont {Tang}, \citenamefont {Hu}, \citenamefont {Ma},\ and\ \citenamefont {Ni}}]{Duality_Lei_et_all_PRL2022}%
  \BibitemOpen
  \bibfield  {author} {\bibinfo {author} {\bibfnamefont {Q.-L.}\ \bibnamefont {Lei}}, \bibinfo {author} {\bibfnamefont {F.}~\bibnamefont {Tang}}, \bibinfo {author} {\bibfnamefont {J.-D.}\ \bibnamefont {Hu}}, \bibinfo {author} {\bibfnamefont {Y.-q.}\ \bibnamefont {Ma}}, \ and\ \bibinfo {author} {\bibfnamefont {R.}~\bibnamefont {Ni}},\ }\href {\doibase 10.1103/PhysRevLett.129.125501} {\bibfield  {journal} {\bibinfo  {journal} {Phys. Rev. Lett.}\ }\textbf {\bibinfo {volume} {129}},\ \bibinfo {pages} {125501} (\bibinfo {year} {2022})}\BibitemShut {NoStop}%
\bibitem [{\citenamefont {Fruchart}\ \emph {et~al.}(2023)\citenamefont {Fruchart}, \citenamefont {Yao},\ and\ \citenamefont {Vitelli}}]{Duality_Fruchart_et_all_PRResearch2023}%
  \BibitemOpen
  \bibfield  {author} {\bibinfo {author} {\bibfnamefont {M.}~\bibnamefont {Fruchart}}, \bibinfo {author} {\bibfnamefont {C.}~\bibnamefont {Yao}}, \ and\ \bibinfo {author} {\bibfnamefont {V.}~\bibnamefont {Vitelli}},\ }\href {\doibase 10.1103/PhysRevResearch.5.023099} {\bibfield  {journal} {\bibinfo  {journal} {Phys. Rev. Res.}\ }\textbf {\bibinfo {volume} {5}},\ \bibinfo {pages} {023099} (\bibinfo {year} {2023})}\BibitemShut {NoStop}%
\bibitem [{\citenamefont {Zhang}\ \emph {et~al.}(2023)\citenamefont {Zhang}, \citenamefont {Lu},\ and\ \citenamefont {Chen}}]{Zhang_et_all_kagome_2023_PhysRevApplied}%
  \BibitemOpen
  \bibfield  {author} {\bibinfo {author} {\bibfnamefont {Z.-D.}\ \bibnamefont {Zhang}}, \bibinfo {author} {\bibfnamefont {M.-H.}\ \bibnamefont {Lu}}, \ and\ \bibinfo {author} {\bibfnamefont {Y.-F.}\ \bibnamefont {Chen}},\ }\href {\doibase 10.1103/PhysRevApplied.20.054002} {\bibfield  {journal} {\bibinfo  {journal} {Phys. Rev. Appl.}\ }\textbf {\bibinfo {volume} {20}},\ \bibinfo {pages} {054002} (\bibinfo {year} {2023})}\BibitemShut {NoStop}%
\bibitem [{\citenamefont {Azizi}\ \emph {et~al.}(2024)\citenamefont {Azizi}, \citenamefont {Sarkar}, \citenamefont {Sun},\ and\ \citenamefont {Gonella}}]{Azizi_2024_DW_PhysRevB}%
  \BibitemOpen
  \bibfield  {author} {\bibinfo {author} {\bibfnamefont {P.}~\bibnamefont {Azizi}}, \bibinfo {author} {\bibfnamefont {S.}~\bibnamefont {Sarkar}}, \bibinfo {author} {\bibfnamefont {K.}~\bibnamefont {Sun}}, \ and\ \bibinfo {author} {\bibfnamefont {S.}~\bibnamefont {Gonella}},\ }\href {\doibase 10.1103/PhysRevB.110.L060102} {\bibfield  {journal} {\bibinfo  {journal} {Phys. Rev. B}\ }\textbf {\bibinfo {volume} {110}},\ \bibinfo {pages} {L060102} (\bibinfo {year} {2024})}\BibitemShut {NoStop}%
\end{thebibliography}%

\onecolumngrid

\section*{\Large\bf Supplemental Material}
\makeatletter

\renewcommand \thesection{S-\@arabic\c@section}
\renewcommand\thetable{S\@arabic\c@table}
\renewcommand{\thefigure}{S\arabic{figure}}
\renewcommand \theequation{S\@arabic\c@equation}
\makeatother
\setcounter{equation}{0}  
\setcounter{figure}{0}  
\setcounter{section}{0}  

\maketitle
\section{Different Mode Shapes of the $180^\circ$ and $0^\circ$ Twisted Kagome Configurations}\label{sec.1}

\begin{figure}[h!]
\centering
\includegraphics[width = 0.6\textwidth]{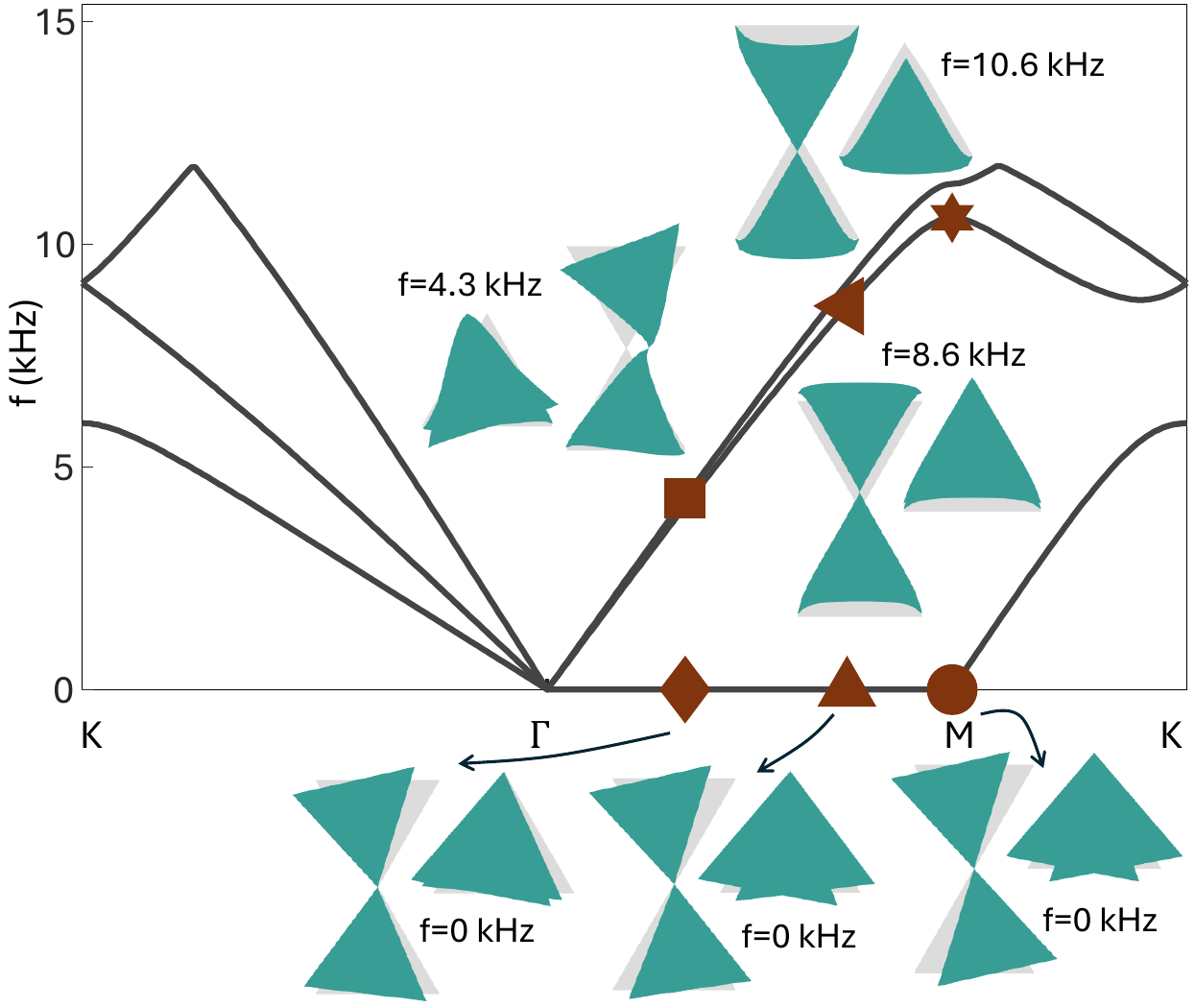}
\caption{Mode shapes of the $180^\circ$ and $0^\circ$ twisted kagome unit cells corresponding to distinct wave vectors along the $\Gamma-M$ direction for the zero-frequency branch and the first finite-frequency acoustic phonon mode.}
\label{FigS1}
\end{figure}

In Fig.1(a) of the main text, we show the band diagrams of some examples of twisted kagome dual-pairs. The two extreme configurations correspond to the deployed ($\theta=180^\circ$) and the compact state ($\theta=0^\circ$), both of which feature bulk floppy modes that manifest as a zero-frequency branch in the $\Gamma-M$ direction. To emphasize the floppy nature of this mode, which is energy-costless, we examine the mode shape samples along that segment. We specifically choose six arbitrary modes associated with distinct wave vectors along the $\Gamma-M$ direction and display their corresponding mode shapes for the two unit cells, as shown in Fig.~\ref{FigS1}. It can be seen that three of the mode shapes having zero frequency involve rigid rotation of the triangles in the unit cells, which costs no elastic energy and thus are defined as bulk floppy modes. In contrast, eigenmodes corresponding to the finite frequency phonon modes display internal deformation of the triangles, involving elastic deformation.

\section{Physical realization of a finite kagome domain incorporated with the z-shift}\label{sec.2}
\begin{figure}[h!]
   \centering
  \includegraphics[width = 0.8\textwidth]{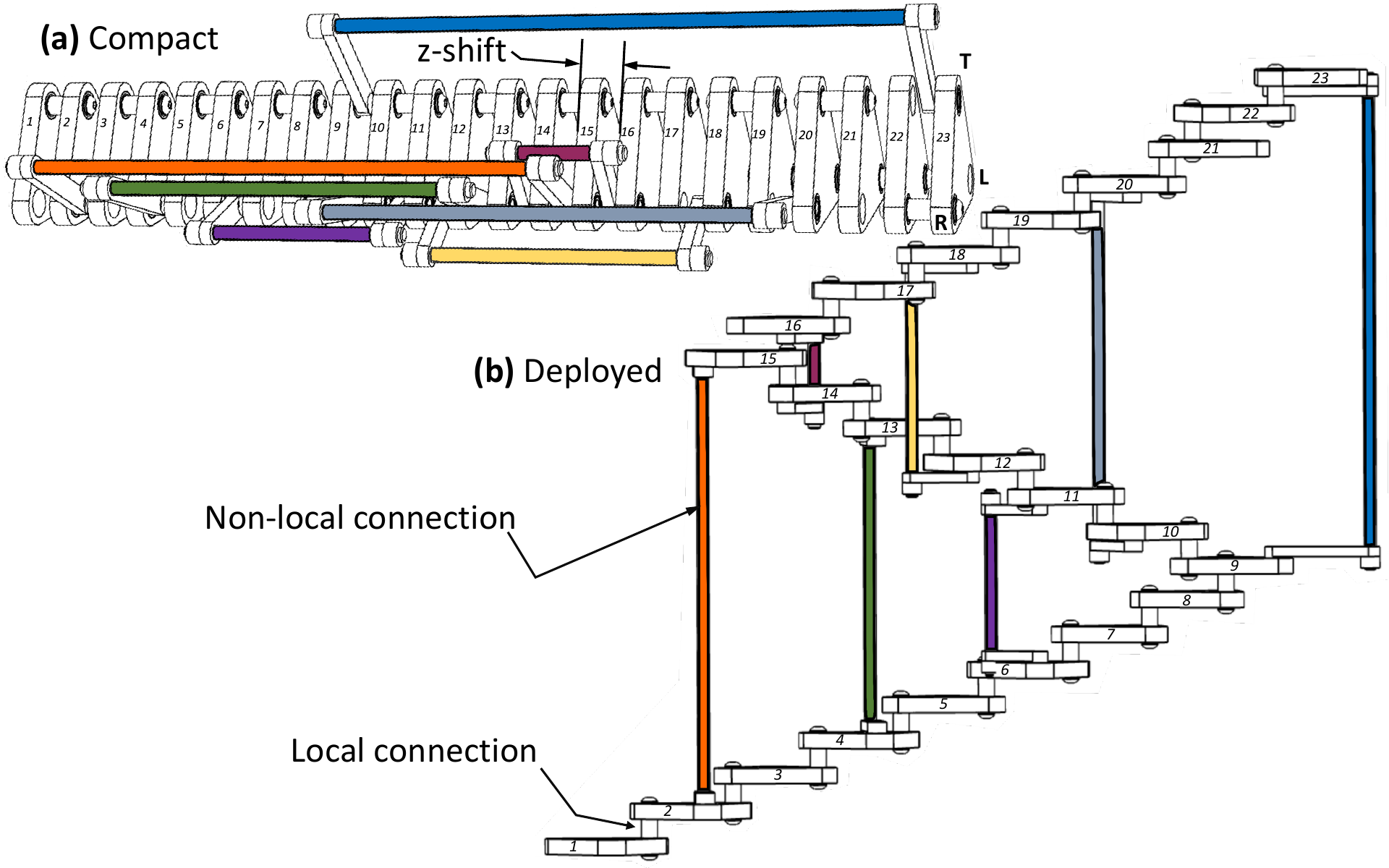}
   \caption{Possible design solutions that implement the lattice shown in Figs.~1(b-c) of the main text via z-shift, shown in the compact (chain) and deployed states, respectively. Non-local interactions are color-coded between the two configurations.}
   \label{FigS2}
\end{figure}

Figs.~\ref{FigS2}(a-b) show the ingredients for a possible physical realization (depicted in the compact and deployed states, respectively) incorporating the required $z$-shift. In both states, we mark the color-coded connections that acquire non-local interactions in the chain. The adopted numbering scheme is just one among the possible options, chosen to minimize the number of non-local connections, thereby simplifying the manufacturing challenges expected in the fabrication of a prototype. This strategy suggests the potential for deploying the lattice in 3D using various patterns, which could expand the scenarios explored in this study and reveal alternative functional landscapes. 

\section{Bulk Floppy Mode manifestation in the deployed lattice}\label{sec.3}

\begin{figure}[h!]
    \centering
    \includegraphics[width=\textwidth]{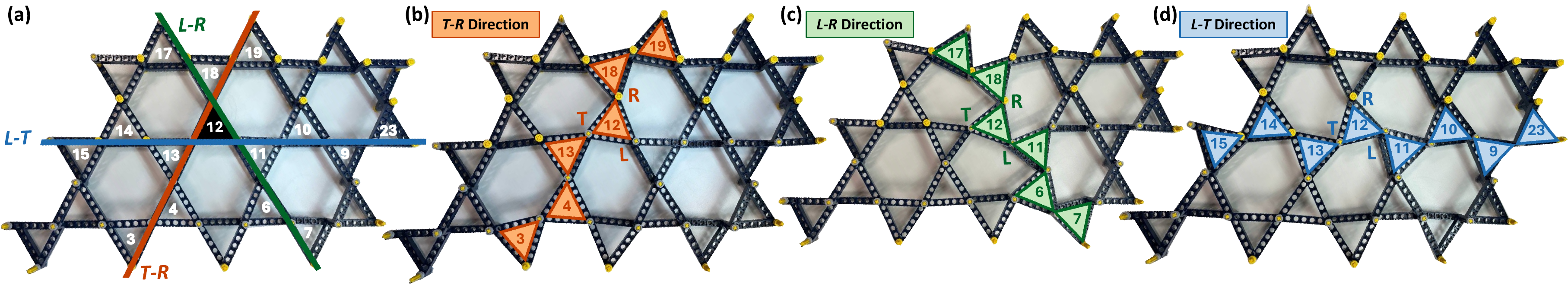}
    \caption{(a) Finite domain kagome prototype constructed with LEGO\textsuperscript{\textregistered} Technic elements, with three straight lines passing through triangle No.~12 marked. (b-d) Snapshots of the lattice behavior when manually loaded at triangle No.~12 along the \textit{T-R}, \textit{L-R}, and \textit{L-T} directions, respectively.}
    \label{FigS3}
\end{figure}

We construct a finite kagome domain prototype composed of 23 triangles, using LEGO\textsuperscript{\textregistered} liftarms connected with nearly frictionless hinges made from LEGO\textsuperscript{\textregistered} axles and bushings. The assembled lattice representing the deployed state is shown in Fig.~\ref{FigS3}(a), with the \textit{T-R}, \textit{L-R}, and \textit{L-T} directions passing through triangle No.~12 highlighted. Figs.~\ref{FigS3}(b-d) demonstrate the lattice behavior when triangle No.~12 is manually loaded statically along its three sides. During each test, applying a force along the desired direction prompts the lattice to develop a floppy mode that localizes along the loading direction. This involves macroscopic rotations of the highlighted triangle sequences in each snapshot along the chosen direction.

\section{Illustration of the Deployment Sequences of the Kagome Chain Prototype}\label{sec.4}
\begin{figure}[h!]
    \centering
    \includegraphics[width=0.5\textwidth]{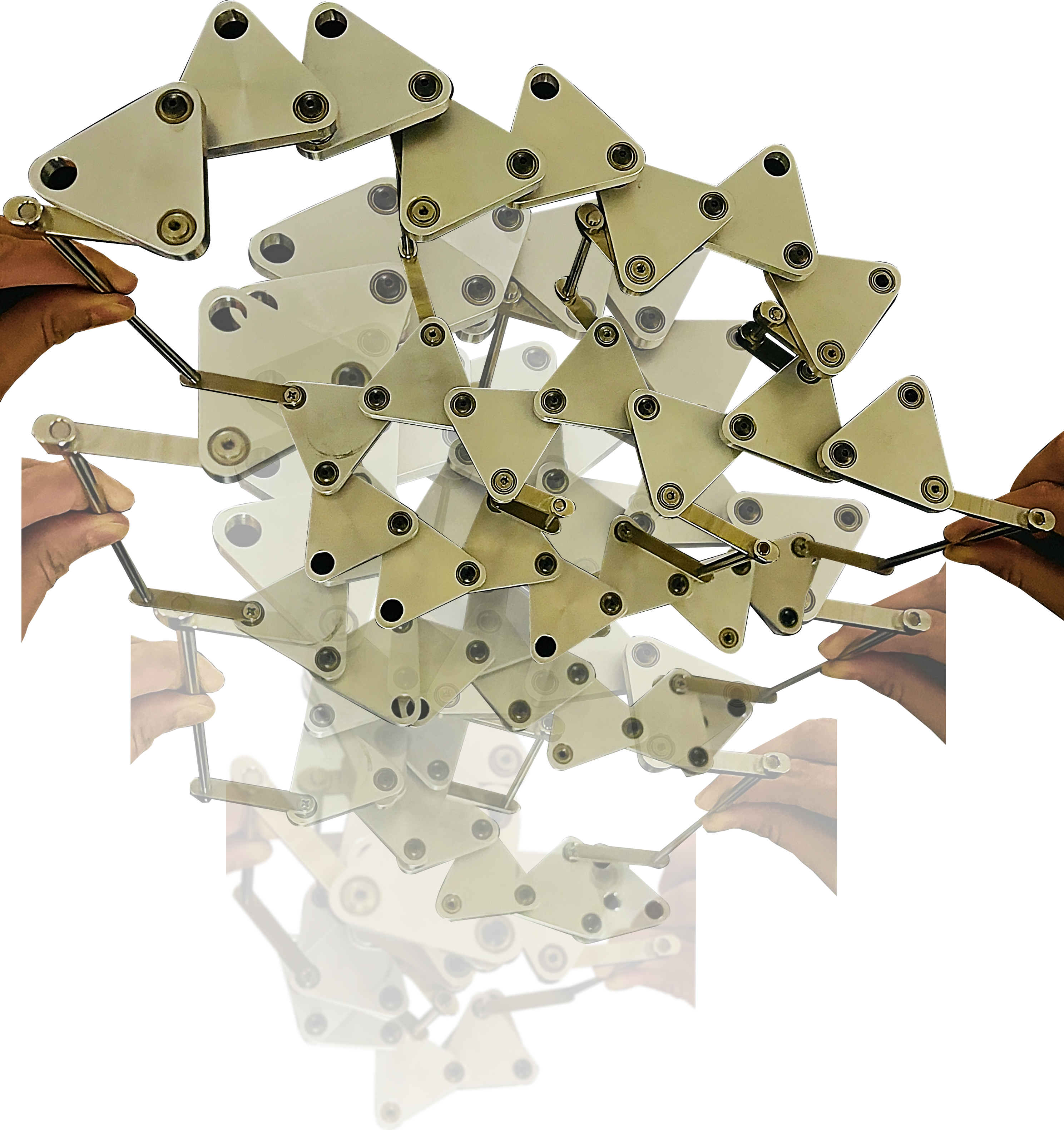}
    \caption{Cascade of frames illustrating the deployment of a twisted kagome lattice endowed with non-local, long-range connections, from the deployed towards the kagome chain state.}
    \label{FigS4}
\end{figure}

Fig.~\ref{FigS4} presents four snapshots of the top views of the chain specimen, showing its progression from the expanded state to the chain state (see supplemental videos for a demonstration of the deployment and retraction process).

\section{SolidWorks Motion Analysis of Activation of Triangles No.~12 and No.~16 in the Chain}\label{sec.5}
\begin{figure}[h!]
    \centering
    \includegraphics[width=0.6\textwidth]{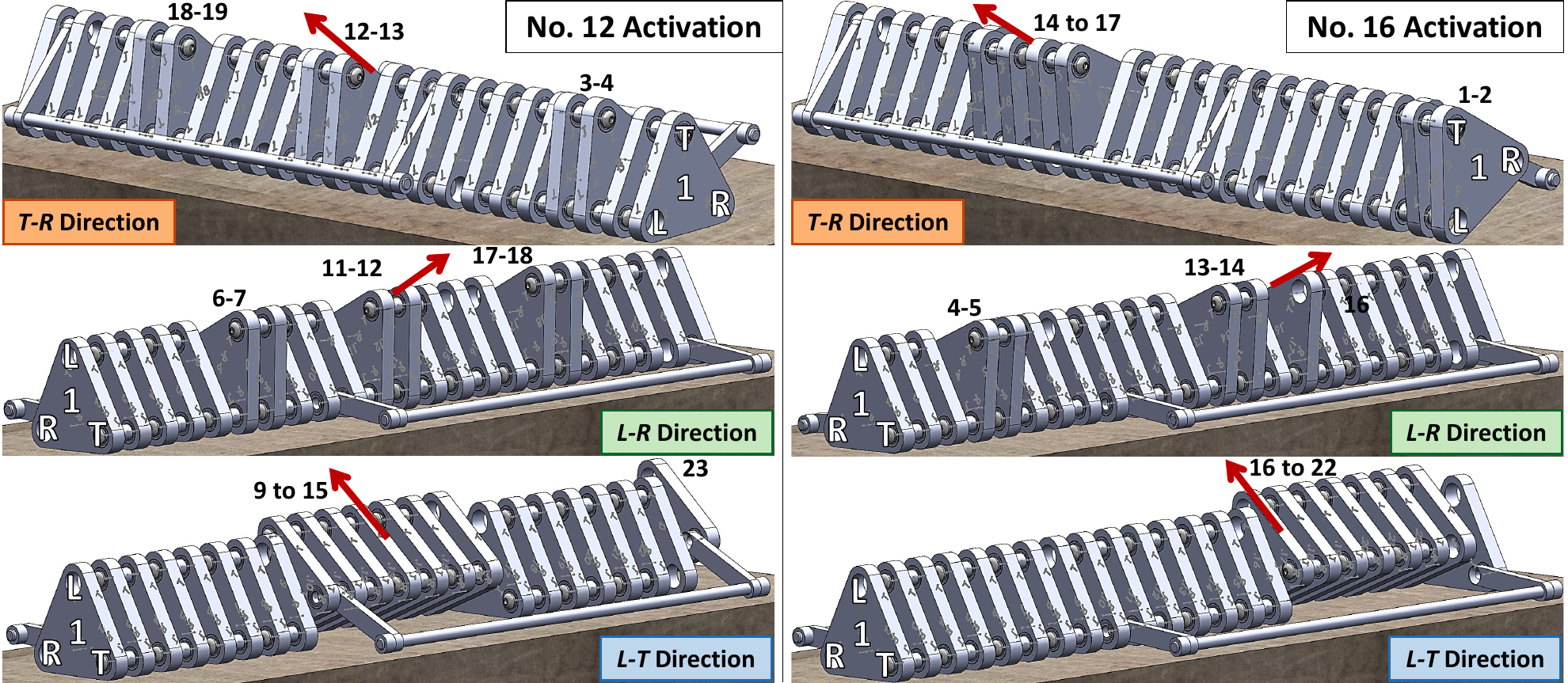}
    \caption{Snapshots of SolidWorks motion analysis of activating triangles No.~12 (left panel) and No.~16 (right panel) along \textit{T-R}, {L-R}, and {L-T} directions. The direction of loading and the rotated triangle sequences are marked in each case. Simulations closely replicate the experimental setup discussed in the main text.}
    \label{FigS5}
\end{figure}
We assemble the kagome chain model in SolidWorks, assigning material properties and defining necessary mates to realistically restrict component motion within the assembly. We then conduct motion studies for the three loading directions of \textit{T-R}, \textit{L-R}, and \textit{L-T}. To simulate realistic conditions, we include a gravitational force and model contacts to account for friction and prevent non-local interactions and the chain from passing through each other or through the floor during motion. We apply a linear, action-only constant force at the desired corner (either \textit{T} or \textit{L}) of the selected triangle (No.~12 or No.~16), ensuring the force remains along the chosen side of the triangle (\textit{T-R}, \textit{L-R}, or \textit{L-T}) during motion. Fig.~\ref{FigS5} presents snapshots of the animated rotated chain, highlighting the role of the new table-top testing setup (see the full animations available in the supplemental videos).

\section{ Full-scale static simulation of the compact state for the three possible loading directions}\label{sec.6}

\begin{figure}[h!]
\centering
\includegraphics[width = 0.8\textwidth]{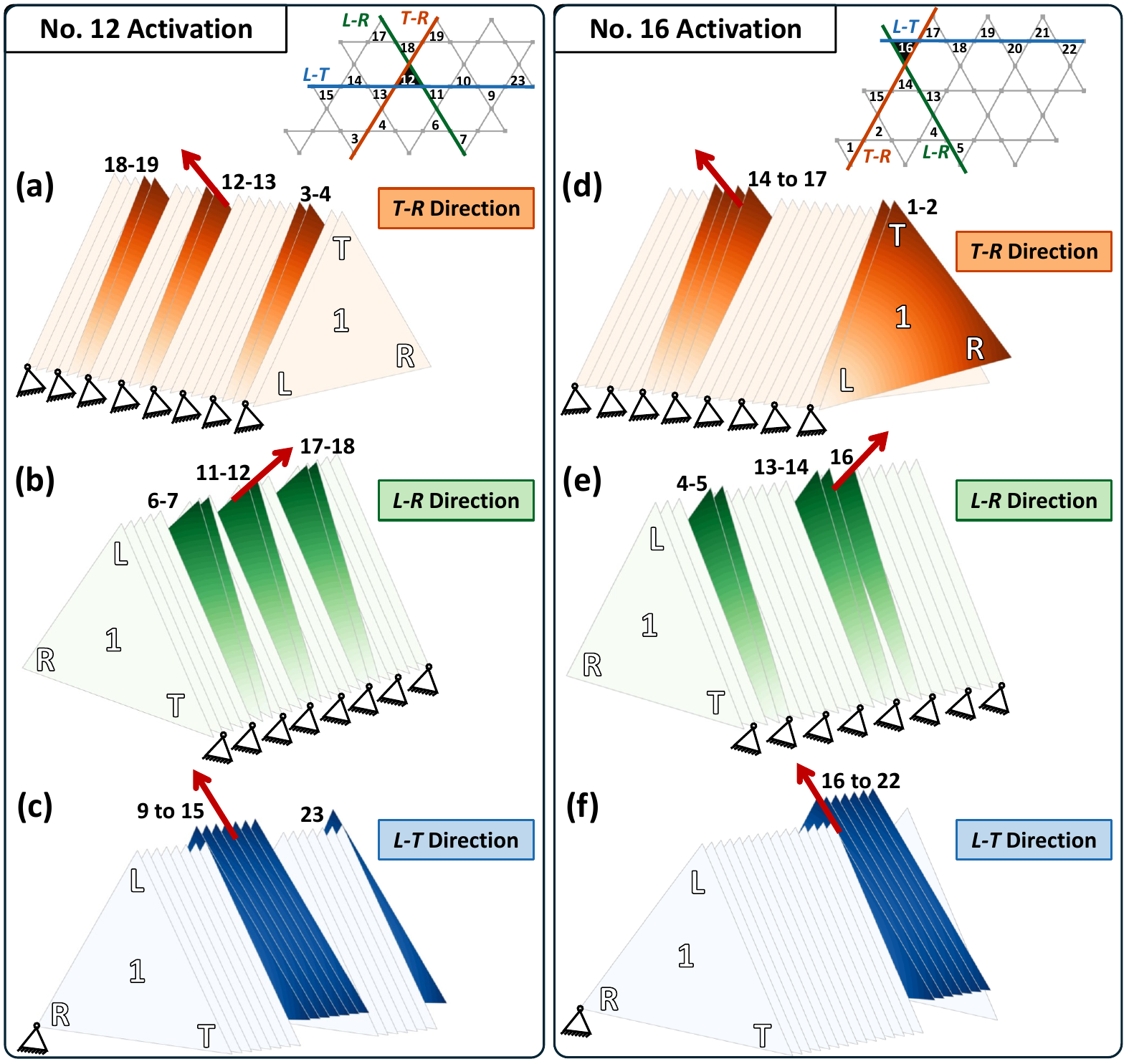}
\caption{Full-scale static simulation of the compact lattice comprising 23 triangles, incorporating updated boundary conditions inspired by the kinematics involved in the table-top chain testing.}
 \label{FigS6}
\end{figure}

Full-scale simulations in the compact state, incorporating the revised boundary conditions associated with the table-top testing of the chain, are shown in Fig.~\ref{FigS6}. We loaded the same triangles (No.~12 and No.~16) along the \textit{T-R}, \textit{L-R}, and \textit{L-T} directions, mimicking the experimental setup kinematics. The activated triangles matched the selected directions in the deployed lattices, a behavior also observed in the experimental testing. We conclude that for each direction of loading, only the triangles located along the loading direction in the deployed state are activated along the chain.

\section{ Full-scale static simulation of the deployed state for the three possible loading directions}\label{sec.7}
\begin{figure}[h!]
   \centering
  \includegraphics[width = \textwidth]{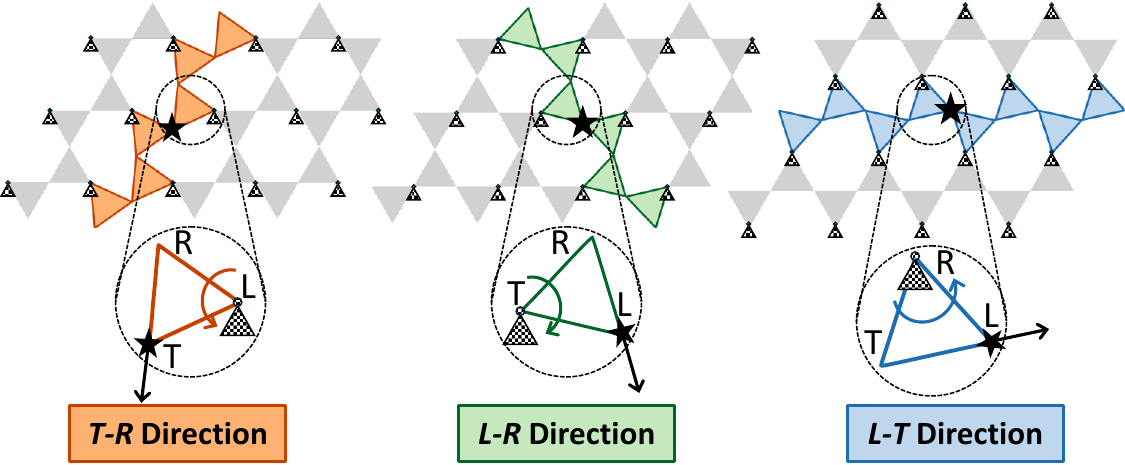}
   \caption{Full-scale static simulation of the fully deployed lattice comprising 23 triangles, incorporating updated boundary conditions inspired by the kinematics involved in the table-top chain testing. The results align with the experiments, supporting the existence of floppy rotations of the expected triangle sequences along the \textit{T-R}, \textit{L-R}, and \textit{L-T} directions.}
   \label{FigS7}
\end{figure}

Full-scale simulations in the deployed state, incorporating the revised boundary conditions associated with the table-top testing of the chain, are shown in Fig.~\ref{FigS7} for the three possible loading directions, along with zoomed-in insets illustrate the applied forces and the resulting moments about. In the absence of net vertical motion and sliding, the loading boils down to a moment about the corner that does not lie along the applied load, which ultimately behaves like a pin support. One can see that soft modes still arise along these three directions, despite the additional constraints.
\end{document}